\documentclass[letterpaper,titlepage,11pt]{article}

\pdfoutput=1

\usepackage{amssymb,amsmath,amsfonts}
\usepackage{epsfig}
\usepackage{ccaption}
\usepackage{graphicx}

\usepackage[
      colorlinks=true,
      linkcolor=blue,
      urlcolor=blue,
      filecolor=black,
      citecolor=red,
      pdfstartview=FitV,
      pdftitle={},
        pdfauthor={Roberto Emparan, Ryotaku Suzuki, Kentaro Tanabe},
        pdfsubject={},
        pdfkeywords={},
        pdfpagemode=None,
        bookmarksopen=true
      ]{hyperref}

\def\clock{{\count0=\time
           \divide\count0 60
           \ifnum\count0<10 0\fi\the\count0
           \multiply\count0 -60 \advance\count0 \time
           :\ifnum\count0<10 0\fi \the\count0
         }}
\newcommand{\timestamp}{{\small\vbox{\hbox{\tt\jobname.tex}
\hbox{\the\day/\the\month/\the\year, \clock}}}}

\newcommand{\nn}{\nonumber}
\newcommand{\ie}{{\it i.e.,\,}}
\newcommand{\eg}{{\it e.g.,\,}}

\newcommand{\lp}{\left(}
\newcommand{\rp}{\right)}

\newcommand{\mc}[1]{\mathcal{#1}}

\newcommand{\beq}{\begin{equation}}
\newcommand{\eeq}{\end{equation}}
\newcommand{\bea}{\begin{eqnarray}}
\newcommand{\eea}{\end{eqnarray}}
\newcommand{\beqa}{\begin{eqnarray}}
\newcommand{\eeqa}{\end{eqnarray}}

\newcommand{\R}{\mathbb{R}}

\newcommand{\sR}{\mathsf{R}}

\numberwithin{equation}{section}
\begin{document}

\begin{titlepage}
%\timestamp
\rightline{KEK-TH-1794} 
\rightline{AP-GR-119}
\rightline{OCU-PHYS-419}
%\leftline{}{\timestamp}
\vskip 1.5cm
\centerline{\LARGE \bf Quasinormal modes of (Anti-)de Sitter black holes}
\medskip
\centerline{\LARGE \bf in the $1/D$ expansion}
\vskip 1.2cm
\centerline{\bf Roberto Emparan$^{a,b}$, Ryotaku Suzuki$^{c}$, Kentaro Tanabe$^{d}$}
\vskip 0.5cm
\centerline{\sl $^{a}$Instituci\'o Catalana de Recerca i Estudis
Avan\c cats (ICREA)}
\centerline{\sl Passeig Llu\'{\i}s Companys 23, E-08010 Barcelona, Spain}
\smallskip
\centerline{\sl $^{b}$Departament de F{\'\i}sica Fonamental, Institut de
Ci\`encies del Cosmos,}
\centerline{\sl  Universitat de
Barcelona, Mart\'{\i} i Franqu\`es 1, E-08028 Barcelona, Spain}
\smallskip
\centerline{\sl $^{c}$Department of Physics, Osaka City University, Osaka 558-8585, Japan}
\smallskip
\centerline{\sl $^{d}$Theory Center, Institute of Particles and Nuclear Studies, KEK,}
\centerline{\sl  Tsukuba, Ibaraki, 305-0801, Japan}
\smallskip
\vskip 0.5cm
\centerline{\small\tt emparan@ub.edu,\, ryotaku@sci.osaka-cu.ac.jp,\, ktanabe@post.kek.jp}

\vskip 1.2cm
\centerline{\bf Abstract} \vskip 0.2cm 
\noindent 
We use the inverse-dimensional expansion to compute analytically the frequencies of a set of quasinormal modes of static black holes of Einstein-(Anti-)de Sitter gravity, including the cases of spherical, planar or hyperbolic horizons. The modes we study are \textit{decoupled} modes localized in the near-horizon region, which are the ones that capture physics peculiar to each black hole (such as their instabilities), and which in large black holes contain hydrodynamic behavior. Our results also give the unstable Gregory-Laflamme frequencies of Ricci-flat black branes to two orders higher in $1/D$ than previous calculations. We discuss the limits on the accuracy of these results due to the asymptotic but not convergent character of the $1/D$ expansion, which is due to the violation of the decoupling condition at finite $D$. Finally, we compare the frequencies for AdS black branes to calculations in the hydrodynamic expansion in powers of the momentum $k$. Our results extend up to $k^9$ for the sound mode and to $k^8$ for the shear mode.

\end{titlepage}
\pagestyle{empty}
\small
%\tableofcontents
\normalsize
\newpage
\pagestyle{plain}
\setcounter{page}{1}

\section{Introduction}

The quasinormal modes of a black hole spacetime encode important aspects of its dynamics \cite{Berti:2009kk}. In particular, the AdS/CFT correspondence implies that the quasinormal modes of Anti-de Sitter black holes describe the relaxation to thermal equilibrium in the dual field theory \cite{Horowitz:1999jd}.
As argued in \cite{Emparan:2014aba} and further elaborated in this article, the limit of large number of dimensions $D$ isolates a subset of quasinormal modes associated to particularly interesting black hole dynamics, and allows efficient analytic computation of their frequencies.

In more detail, when $D$ is very large the gravitational field of a black hole gets strongly localized close to the horizon \cite{Asnin:2007rw,Emparan:2013moa}. The existence of a well-defined `near-horizon region' \cite{Emparan:2013xia} splits the quasinormal spectrum into two distinct sets that capture very different black hole dynamics. The most numerous set ($\propto D^2$) are `non-decoupling' modes that straddle between the near-horizon zone and the asymptotic zone; they are largely insensitive to the peculiarities of the black hole, which for these modes is simply a hole in a background spacetime \cite{Emparan:2014cia}. In contrast, the much smaller set ($\propto D^0$) of `decoupled' quasinormal modes have support only in the near-horizon region, where they are normalizable states to all orders in $1/D$; they capture properties specific to each black hole, such as the instabilities of certain higher-dimensional black holes and black branes, and the hydrodynamic modes of black branes \cite{Emparan:2013moa,Emparan:2014jca}.\footnote{The decoupled spectrum at large $D$ was first identified numerically in \cite{Dias:2014eua}.} 

Due to the simple form of the near-horizon geometry, the decoupled quasinormal frequencies can be calculated perturbatively in analytic form to several orders in $1/D$. Furthermore, the universality of the leading-order near-horizon geometry implies that the structure of the calculation is essentially the same for all static neutral black holes, be they asymptotic to de Sitter, Minkowski, or Anti-de Sitter, with spherical, flat or hyperbolic horizons. This allows us to obtain unified analytical formulas for their frequencies, valid up to next-to-next-to-next-to-leading order (3NLO) in $1/D$, and for planar horizons up to 4NLO. Via the AdS/Ricci-flat correspondence \cite{Caldarelli:2012hy}, the latter also give the unstable Gregory-Laflamme (GL) frequencies of Ricci-flat black branes \cite{Gregory:1993vy} to two higher orders in $1/D$ than in \cite{Emparan:2013moa}.

For AdS black branes a hydrodynamic gradient expansion has been applied to obtain the sound- and shear-mode dispersion relations $\omega(k)$ analytically (exactly in $D$) up to $k^3$ for the sound mode, and up to $k^4$ for the shear mode \cite{Bhattacharyya:2008mz}. Where there is overlap, our calculations perfectly match these results. As we will see, the regime of applicability for the large $D$ expansion is not essentially different than for the hydrodynamic expansion. However, the large $D$ expansion allows a much simpler calculation of  higher orders in $k$: our results extend up to $k^9$ for the sound mode, and up to $k^8$ for the shear mode.

It is natural to ask whether these higher-order corrections always increase the accuracy of the results when finite, possibly low, values of $D$ are substituted. In other words, is the $1/D$ expansion for the decoupled quasinormal spectrum a convergent series? We will explain that it is not, but is instead only asymptotic. The decoupling condition, which holds to all perturbative orders, is violated by effects that are not perturbative in $1/D$. Nevertheless, we obtain excellent agreement for certain quantities, such as the GL critical wavelength, even down to $D=6$, which we reproduce within $\sim 2\%$ accuracy.

In the next section we introduce the class of black holes we study and their large $D$ limit. In sec.~\ref{sec:qual} we explain qualitatively the appearance of the decoupled spectrum and its main features. The main results, namely the scalar and vector decoupled frequency spectra and the GL unstable frequencies, are given in sec.~\ref{sec:quant}. Sec.~\ref{sec:asym} examines the lack of convergence of the $1/D$ series for decoupled modes. Sec.~\ref{sec:hydro} compares our results with those obtained exactly in $D$ in the hydrodynamic expansion at low momenta. We conclude in sec.~\ref{sec:conc}. Appendices \ref{app:scpert} to \ref{app:adsbh} contain further technical details and lengthy results. In particular, appendix \ref{app:adsbh} gives explicit expressions for the frequencies for spherical horizons.

\section{Set up}

We consider (A)dS black holes with metric
%============<Equation>=============%
%
\begin{eqnarray}\label{adsbh}
ds^{2}=-f(r)dt^{2}+f(r)^{-1}dr^{2}+r^{2}\,d\sigma^{2}_{K,D-2},
\end{eqnarray}
%
%==================================%
with
%============<Equation>=============%
%
\beq
f(r)=
K-\lambda\frac{r^{2}}{r_0^2}-\left( \frac{r_{0}}{r} \right)^{D-3} \,,
\eeq
and where 
\beq
\lambda=\frac{2\Lambda r_0^2}{(D-1)(D-2)}
\eeq
parametrizes the cosmological constant $\Lambda$ in units of the black hole size-scale $r_0$. 
The solution has another, discrete parameter,
\beq
K=\pm 1,0
\eeq
for the curvature of the
metric $d\sigma^{2}_{K,D-2}$ on the `orbital' space $\mathcal{K}_{D-2}$: a unit sphere, a plane, or a hyperboloid, for $K=+1,0,-1$ respectively.

It is convenient to use 
\beq\label{nD3}
n=D-3
\eeq
instead of $D$ as the perturbation parameter, and
\beq
\sR=\lp \frac{r}{r_0}\rp^n
\eeq
as the radial variable appropriate for the near-horizon region, which is defined by $\sR\ll e^n$. In terms of it, the metric \eqref{adsbh} is
\beq
ds^2=-f(\sR)dt^2+\frac{r_0^2}{n^2}\sR^{2/n}\frac{d\sR^2}{\sR^2 f(\sR)}+\sR^{2/n}r_0^2\,d\sigma^{2}_{K,D-2}
\eeq
with
\beq
f(\sR)=K-\lambda \sR^{2/n} -\frac1{\sR}\,.
\eeq
The horizon at $\sR=\sR_H$ such that $f(\sR_H)=0$, has surface gravity $\kappa$ given by
\beq
\kappa r_0=\frac{nK}{2\sR_H^{1/n}}-\lp\frac{n}{2}+1\rp \lambda \sR_H^{1/n}\,.
\eeq

Taking $n\to\infty$ with $\sR$ and $\lambda$ finite, to leading order the geometry in the $(t,\sR)$ directions is\footnote{We omit here the $\mathcal{K}_{D-2}$, whose size plays the role of the dilaton for the effective two-dimensional gravity \cite{Emparan:2013xia}.} 
\beq\label{nearh}
ds^2_{(2d)}=-\lp \frac1{\sR_0}-\frac1{\sR}\rp dt^2+\frac{r_0^2}{n^2}\frac{d\sR^2}{\sR^2}\lp \frac1{\sR_0}-\frac1{\sR}\rp^{-1}
\eeq
with 
\beq
\sR_0=\frac1{K-\lambda}\,.
\eeq
If we rescale $(t,\sR)\to ( \sR_0 t, \sR_0 \sR)$ the metric becomes the same for all values of $K$ and $\lambda$. However, subleading corrections distinguish among them, as we can see in the horizon location which is at
\beq
\sR_H=\sR_0+\frac{2\lambda \sR_0^2 \ln \sR_0}{n}+\mc{O}\lp n^{-2}\rp\,.
%+\frac{2\lambda \sR_0^3 \ln \sR_0 \lp 2\lambda +(2\lambda+\sR_0^{-1})\ln \sR_0\rp}{n^2}+\mc{O}\lp \frac1{n^3}\rp\,.
\eeq

In the planar case $K=0$ all values of $\lambda<0$ are equivalent to $\lambda=-1$. In the hyperbolic case $(K=-1, \lambda<0$), when $n$ is finite there exist black hole solutions with negative mass parameter $r_0^n<0$, with the lowest negative mass corresponding to an extremal black hole. It is easy to see, however, that as $n\to\infty$ the range of allowed negative masses shrinks to zero, and as a consequence $\lambda$ is bounded above by $\lambda< -1$. 
More generally, the large $n$ limit yields a good near-horizon geometry only when
\beq\label{Klam}
\sR_0>0\,.
\eeq
Note that the hyperbolic slicing of AdS (obtained for $r_0^n=0$) lies outside our study, which is expected since there is no strong localization of the field close to an acceleration horizon.

\section{Decoupled spectrum: qualitative aspects}\label{sec:qual}

Before entering the calculational details, let us explain the existence and basic properties of the decoupled quasinormal spectrum at large $D$. 

At first sight, one would not expect any decoupled dynamics in the near-horizon region of neutral static black holes: this region has only a very short radial extent $\sim r_0/D$ away from the horizon, in contrast to the long throats of (near-)extremal black holes which support excitations that can propagate inside them without ever leaving. Indeed, the excitations with frequencies $\omega_\mathrm{nh}\sim D/r_0$ characteristic of the near zone are not decoupled: they are non-normalizable states of the near-horizon geometry \cite{Emparan:2014aba}. However, as we will see presently, this geometry supports a few modes that, to leading order, are \textit{not dynamical}: in near-horizon scales they are zero-modes with frequency $\sim \omega_\mathrm{nh}/D\to 0$. Furthermore, the universality of the leading-order near-horizon geometry implies that these modes are present in all neutral black holes. 

\begin{figure}[t]
 \begin{center}
  \includegraphics[width=.44\textwidth,angle=0]{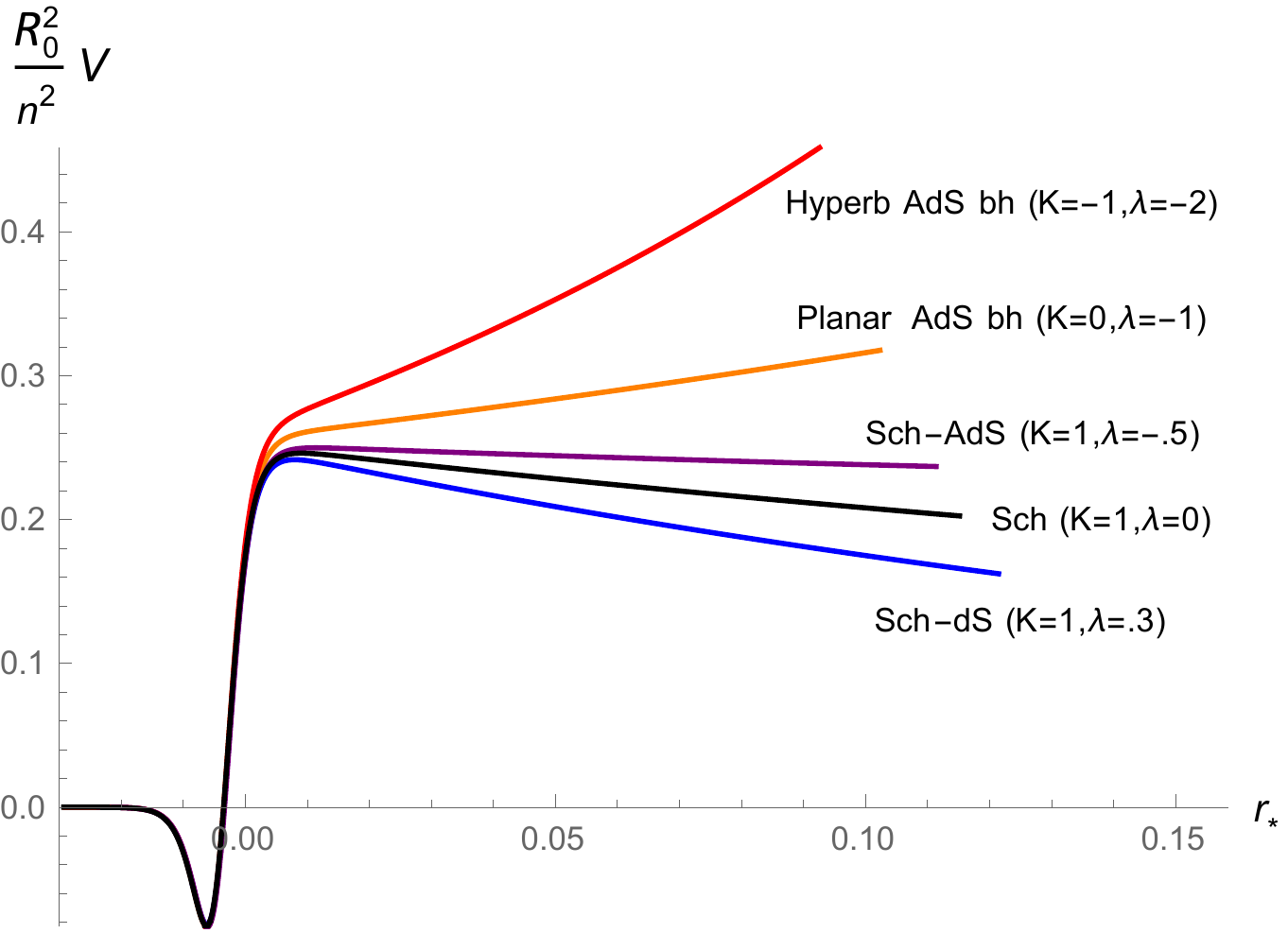}
  \hspace{5mm}
  \includegraphics[width=.51\textwidth,angle=0]{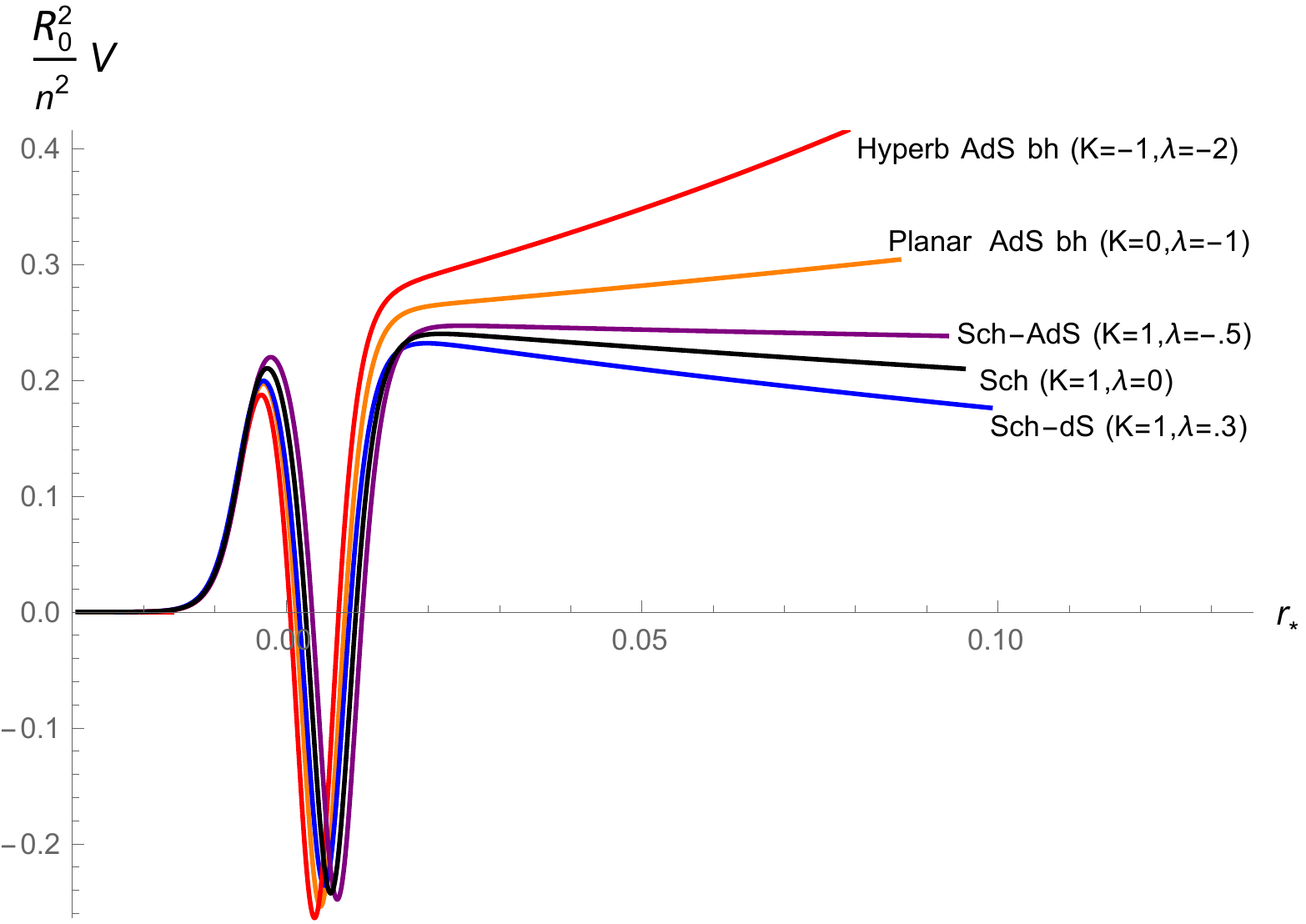}
   \end{center}
% \vspace{-5mm}
 \caption{\small Radial potentials $V(r_*)$ of gravitational vector (left) and scalar (right) perturbations, for $n=500$ and harmonic eigenvalue $q^2=2 n$, for representative examples of (A)dS black holes. The plots show the potential around the region where the near and far zones meet (the horizon is at $r_*\to-\infty$). Decoupled quasinormal modes are localized around the minima in the near-zone. The potentials for tensor perturbations differ only in the absence of these minima.}
 \label{fig:potentials}
\end{figure}

Such leading-order zero-modes exist for gravitational perturbations which are vectors or scalars of the space $\mathcal{K}_{D-2}$, and are absent for gravitational tensors (and also for free massless scalar fields, which satisfy the same equation). To see this point, note that the wave equation can in all cases be written in the form \cite{Kodama:2003jz}
\beq\label{waveeq}
\lp \frac{d^2}{d r_*^2}+\omega^2-V_s(r_*)\rp \Psi_s (r_*)=0
\eeq
where $r_*=\int dr/f(r)$ is the conventional tortoise coordinate and \beq
s=0,1,2
\eeq 
denote gravitational scalars, vectors and tensors, respectively. The explicit form of the potentials $V_s$ will be given later below; their form at large $D$ is presented in fig.~\ref{fig:potentials} for vectors and scalars, zooming in around the most interesting region, $r_*=\mc{O}(1)$, where the near and far zones meet. At large $r_*$ the potentials reflect the properties of the background: decaying for asymptotically dS and flat space, and growing (`bounding box') for asymptotically AdS. In contrast, near the horizon the form of the potential becomes the same independently of the cosmological constant or the horizon curvature. In the direction of decreasing $r_*$, as we enter the near-horizon zone the potential drops exponentially from a height $\sim D^2/r_0$.
In this near-horizon region the potentials for gravitational scalars and vectors have negative minima, which are absent for tensors.\footnote{Fig.~\ref{fig:potentials} leads one to expect that the $1/n$ expansion is more accurate for vectors than for scalars, which indeed appears to be the case.} These minima allow the existence of non-trivial solutions to the equation \eqref{waveeq} with $\omega=0$. Since at large $D$ both $d^2/d r_*^2$  and $V_s$ scale like $D^2/r_0^2$, these zero-modes are actually leading-order solutions to \eqref{waveeq} for modes with frequency $\omega =o(D/r_0)$.

These leading-order static modes become dynamical, \ie\ their frequency is non-zero $\omega\sim 1/r_0$, at the next order in the $1/D$ expansion. At this order, the differences between black holes (due to the cosmological constant or the horizon curvature) make themselves present and therefore the frequencies differ for each of them. However, the structure of the perturbation expansion is the same for all of them since it is controlled by the leading-order solution.

\section{Decoupled spectrum: analytic results}\label{sec:quant}

Henceforth in this section we use units where 
\beq
r_0=1\,.
\eeq 
The explicit form of the potentials $V_s$ is \cite{Kodama:2003jz}
%============<Equation>=============%
%
\beq
V_{s=2} = \frac{f(\sR)}{\sR^{2/n}}\lp
q^{2} +2K +\frac{n(n+1)}{2}\sR \frac{df(\sR)}{d\sR} +\frac{n^2-1}{4}f(\sR)
\rp, 
\eeq
%
%==================================%
%============<Equation>=============%
%
\beq
V_{s=1} = \frac{f(\sR)}{\sR^{2/n}}\lp
q^{2} + \frac{n^{2}+3}{4}K 
 -\frac{3(n+1)^{2}}{4}\frac1{\sR}-\frac{n^2-1}{4} \lambda \sR^{2/n}
\rp\,,
\eeq
%
%==================================%
and
%============<Equation>=============%
%
\begin{eqnarray}
&&
V_{s=0} = \frac{f(\sR)}{\sR^{2/n}}\frac{Q(\sR)}{\lp 4m+\frac{2(n+1)(n+2)}{\sR}\rp^{2}},
\end{eqnarray}
%
%==================================%
where $m=q^{2}-(D-2)K$ and 
%============<Equation>=============%
%
\begin{eqnarray}
Q(\sR) &=& \frac{(n+1)^{4}(n+2)^{2}}{\sR^{3}} \notag\\ 
&& + \frac{(n+1)(n+2)}{\sR^2}
\lp 4m(2n^2+n+3) +(n+2)(n-3)(n^2-1)K \rp \notag \\
&&
-\frac{12(n+1)m}{\sR}\lp (n-3)m +(n+2)(n^2-1)K \rp \\
&&
 + 4(n+3)(n+1)Km^{2}+16m^{3}\notag \\
&&-\lp \frac{(n+3)(n+2)^2(n+1)^3}{\sR^{2}} -\frac{12(n+3)(n+2)(n+1)m}{\sR} 
+4(n-1)(n-3)m^{2} \rp \lambda \sR^{2/n}\,.\notag
\end{eqnarray}

We use $q^2$ to denote the eigenvalues of the Laplacian on $\mc{K}_{n+1}$. 
It is easy to see that the minima of the potentials and hence the decoupled modes only exist for $s=0,1$ and for eigenvalues such that $q^2/n^2\to 0$ \cite{Emparan:2014aba}. Thus we take $q^2=\mc{O}(n)$ and introduce finite renormalized eigenvalues
\beq
\hat q^2=\frac{q^2}{n}\,.
\eeq
In particular, 
%============<Equation>=============%
%
\begin{eqnarray}\label{eigen}
\hat q^{2} = 
\begin{cases}
~~\ell\lp 1+\frac{\ell}{n}\rp-\frac{s}{n} & (K=1)\,,\\
~~\frac{k^{2}}{n}  & (K=0)\,,
\end{cases}
\end{eqnarray}
%
%==================================%
where $\ell=\mc{O}(n^0)$ is the angular momentum number on $S^{n+1}$ and  $k=\mc{O}(\sqrt{n})$ is the momentum along $\R^{n+1}$.
For $K=-1$, if the hyperboloid is not compactified the spectrum is continuous like in the planar case, but we will not be more specific about this.

Quasinormal modes are solutions to \eqref{waveeq} that satisfy specific boundary conditions.
For decoupled modes, these are as follows. At large distance in the near-horizon geometry, where $1\ll \sR\ll e^n$, we impose
\beq\label{decoup}
\Psi_s(\sR\to\infty)\to \frac1{\sqrt{\sR}}
\eeq
at all orders in the $1/n$ expansion. This expresses that the mode is normalizable in this geometry (non-normalizable modes are $\propto\sqrt{\sR}$ at large $\sR$). Since it is (perturbatively) shielded from the far zone, it can be matched to a purely outgoing wave in a Minkowski background \cite{Emparan:2014aba} or to other suitable function for other asymptotics.  

The solution must be ingoing at the future horizon. This is achieved when
\beq
\Psi_s(\sR)= \lp \sR-\sR_H\rp^{-i\omega/(2\kappa)} \phi_s(\sR)
\eeq
where $\phi_s(\sR)$ is regular at $\sR=\sR_H$. 

Observe that 
\begin{itemize}
\item the condition at $\sR\to\infty$ is independent of the asymptotics in the far zone; \item since $\kappa\propto n$, the horizon condition to leading order in $1/n$ is simply finiteness of $\Psi_s(\sR_0)$. 
\end{itemize}
These two properties, together with the universal character of the geometry \eqref{nearh}, imply that the leading order solutions (the zero modes) are the same for all values of $K$ and $\lambda$, which allows to study these quasinormal modes in a unified manner for all the metrics \eqref{adsbh}.

The perturbative calculation of the decoupled quasinormal frequencies now proceeds as for Schwarzschild black holes $(K=1,\lambda=0,\sR_0=1)$ in ref.~\cite{Emparan:2014aba}. Like in that instance, for the scalar perturbations it is more convenient to use a formulation in other variables than $\Psi_{s=0}(\sR)$, which we give in appendix~\ref{app:scpert}. We omit the lengthy but straightforward details of the calculations and just give the final results. Writing
\beq
\omega=\sum_{i\geq 0}\frac{\omega_{(i)}}{n^i}
\eeq
we obtain
\subparagraph{Vector-type quasinormal frequencies:}
\beqa
\omega_{(0)}&=&i \lp K-\hat{q}^2\rp ,\\
\omega_{(1)}&=&-i \lp K-\hat{q}^2\rp  \lp \ln \sR_0+2\rp ,\\
\omega_{(2)}&=&-\frac{i}{6}  \lp K-\hat{q}^2\rp  \lp 2\sR_0 \lp \lambda  \lp 6 \ln \sR_0+\pi ^2\rp -\pi ^2 \hat{q}^2\rp -3 \ln \sR_0 \lp \ln \sR_0+4\rp +2 \lp \pi ^2-12\rp \rp ,\\
\omega_{(3)}&=&-\frac{i}{6} \lp K-\hat{q}^2\right) \Biggl[ 
\lp \ln \sR_0   \lp \ln \sR_0  +6\rp  -2 \pi ^2+24\rp  \ln \sR_0 -8 \lp \pi ^2-6\rp\notag\\
&&\qquad\qquad\qquad  
+2\sR_0 \biggl( 12\sR_0 \zeta (3) \lp \lambda -\hat{q}^2\rp  \lp K-\hat{q}^2\rp +\pi ^2 \hat{q}^2 \lp \ln \sR_0  +4 -2 \lambda \sR_0 \lp \ln \sR_0  +2\rp \rp \notag\\
&&\qquad\qquad\qquad\qquad\qquad +2 \lambda ^2\sR_0 \lp \ln \sR_0  +2\rp  \lp 3 \ln \sR_0  +\pi ^2\rp +\lp \pi ^2-12\rp  \lambda  \ln \sR_0  \biggr)\Biggr] \,.
\eeqa
Vector modes are purely imaginary. For $K=1$ a perturbation with $\hat q^2=1$, \ie\ $\ell=1$, is not a quasinormal mode but corresponds to adding angular momentum to the black hole.

\subparagraph{Scalar-type quasinormal frequencies:}
\beqa
\textrm{Re}\,\omega_{(0)}&=&  \sqrt{\frac{\hat{q}^2}{\sR_0} -K^2}\,,\\
\textrm{Re}\,\omega_{(1)}&=&   \frac{\left(K-\hat{q}^2\right) \left(K \left(\ln\sR_0+2\right)-\hat{q}^2\right)-\frac{\hat{q}^2}{2 \sR_0}}{\sqrt{\frac{\hat{q}^2}{\sR_0}-K^2}}
\,,\\
\textrm{Re}\,\omega_{(2)}&=& -\frac{1}{24 \left(\frac{\hat{q}^2}{\sR_0}-K^2\right)^{3/2}}\Biggl[8 K^3 \sR_0 \left(K-\hat{q}^2\right) \left(2 K \left(3 \ln\sR_0+\pi ^2\right)-\pi ^2 \hat{q}^2\right)\nn\\
&&
\qquad\qquad\qquad \qquad\quad -4 \biggl( 3 K^4 \left(\ln\sR_0 \left(\ln\sR_0+8\right)+8\right)+6 K^3 \hat{q}^2 \left(-2 \ln\sR_0+\pi ^2-7\right)\nn\\
	&&\qquad\qquad\qquad\qquad\qquad\quad -K^2 \hat{q}^4 \left(3 \ln\sR_0 \left(\ln\sR_0+8\right)+8 \pi ^2+3\right)\nn\\
	&&\qquad\qquad\qquad\qquad\qquad\quad +2 K \hat{q}^6 \left(3 \ln\sR_0+\pi ^2+9\right)-3 \hat{q}^8\biggr)\nn\\
&&
\qquad\qquad\qquad \qquad\quad+\frac{4 \hat{q}^2}{\sR_0} \biggl(3 K^2 \left(\ln\sR_0 \left(2 \ln\sR_0+11\right)+11\right)\nn\\
	&&\qquad\qquad\qquad\qquad\qquad\qquad +K \hat{q}^2 \left(-3 \ln\sR_0 \left(2 \ln\sR_0+17\right)+2 \pi ^2-69\right)
\nn\\
	&&\qquad\qquad\qquad\qquad\qquad\qquad +\hat{q}^4 \left(12 \ln\sR_0-2 \pi ^2+33\right)\biggr)-\frac{9 \hat{q}^4}{\sR_0^2}\Biggr]
\,,
\eeqa
\beqa
\textrm{Im}\,\omega_{(0)}&=&   K- \hat{q}^2\,,\\
\textrm{Im}\,\omega_{(1)}&=& \hat{q}^2 \left(\ln\sR_0+3\right)-K \left(\ln\sR_0+2\right) \,,\\
\textrm{Im}\,\omega_{(2)}&=& \frac{\sR_0 }{3} \left(K-\hat{q}^2\right) \left(\pi ^2 \hat{q}^2-2 K \left(3 \ln\sR_0+\pi ^2\right)\right)+\frac{K}{2} \left(\ln\sR_0 \left(\ln\sR_0+8\right)+8\right)\nn\\
&& -\frac{\hat{q}^2}{2}  \left(\ln\sR_0 \left(\ln\sR_0+10\right)+14\right)\,.
\eeqa
The expressions for $\omega_{(3)}$ are very lengthy and we defer them to appendix~\ref{app:om3sc}.
The $K=1$ scalar modes with $\ell=1$ are gauge modes.

These results are valid for all $K$ and $\lambda$ satisfying \eqref{Klam}. 
For ease of reference, we give the expressions for the particular case of $K=1$, \ie\ (A)dS Schwarzschild black holes in appendix~\ref{app:adsbh}. For $\lambda=0$ Schwarzschild black holes, these results reproduce those of \cite{Emparan:2014aba}.

\paragraph{Large black hole limit vs.\ large $n$ limit.} At any finite $n$, the large black hole limit $-\lambda\equiv r_0^2/L^2\to \infty$ of the AdS Schwarzschild solution, with $K=1$, results in the AdS black brane. However, when we compute modes of AdS Schwarzschild black holes in the $1/n$ expansion we assume that $r_0/L$ remains finite as $n\to\infty$ (more precisely, we require that $r_0^2/L^2\ll e^n$). Thus the two limits $n\to\infty$ and $r_0/L\to\infty$ need not commute. Indeed, while the leading order terms $\omega_{(0)}$ for AdS black branes \textit{are} correctly obtained as the limit $r_0/L\to\infty$ of AdS Schwarzschild frequencies, the next-to-leading order corrections are not. 

\subsection{AdS black branes}\label{sec:adsb} 

For planar horizons it is ¡possible to perform the analysis in a different manner \cite{Kovtun:2005ev} (see appendix~\ref{app:KS}): given a momentum vector $k^a$ along $\mathbb{R}^{n-1}$, one can decompose the perturbations into scalars, vectors and tensors of its little group $SO(n)$. This has allowed us to carry out the expansion to one higher order than the previous results, up to $1/n^4$. 

For this case, the expansion parameter is more appropriately $1/(D-1)=1/(n+2)$ rather than $1/n$. So here we introduce
\beq
\bar{n}=D-1=n+2\,,
\eeq
and
\beq
\hat{k}^2=\frac{k^2}{\bar{n}}=\hat{q}^2\lp 1-\frac{2}{\bar{n}}\rp\,,
\eeq
and express results as an expansion in $1/\bar{n}$. We find:

\subparagraph{Vector (shear) mode:} 
%============<Equation>=============%
%
\begin{eqnarray}\label{vbb}
\omega &=& -i \hat{k}^{2} \Bigl[ \,
1 + \frac{\pi^{2}\hat{k}^{2}}{3\bar{n}^{2}} - \frac{4\hat{k}^{2}(1+\hat{k}^{2})\zeta(3)}{\bar{n}^{3}} +\frac{4\pi^{4}\hat{k}^{2}(1+7\hat{k}^{2}+\hat{k}^{4})}{45\bar{n}^{4}} +\mc{O}(\bar{n}^{-5})
\Bigr], \notag \\
&=& 
-i \hat{k}^{2} \Bigl[ \,
1 + 2\zeta(2)\frac{\hat{k}^{2}}{\bar{n}^{2}} - 4\zeta(3)\frac{\hat{k}^{2}+\hat{k}^{4}}{\bar{n}^{3}} %\notag \\
+8\zeta(4)\frac{\hat{k}^{2}+7\hat{k}^{4}+\hat{k}^{6}}{\bar{n}^{4}} +\mc{O}(\bar{n}^{-5})
\,\Bigr].
\end{eqnarray}
%
%==================================%

\subparagraph{Scalar (sound) mode:}

%============<Equation>=============%
%
\begin{eqnarray}\label{Rescbb}
{\rm Re}\, \omega_{\pm}&=&\pm\hat{k}\Biggl[
1+\frac{1+2\hat{k}^{2}}{2\bar{n}}
+\frac{1}{\bar{n}^{2}}\left( \frac{3}{8}-\frac{\hat{k}^{2}}{2}+\frac{\pi^{2}\hat{k}^{2}}{3}-\frac{\hat{k}^{4}}{2} \right)\notag \\
&&
+\frac{1}{\bar{n}^{3}}\left( \frac{5}{16} -\hat{k}^{2}\left( \frac{9}{8}+\frac{\pi^{2}}{6}+4\zeta(3) \right) 
+\hat{k}^{4}\left( \frac{3}{4}+\pi^{2}-2\zeta(3) \right) +\frac{\hat{k}^{6}}{2}\right) \notag \\
&&
+\frac{1}{\bar{n}^{4}} \Biggl(
\frac{35}{128} +\hat{k}^{2}\left(-\frac{25}{16} -\frac{3\pi^{2}}{8}+\frac{4\pi^{4}}{45}+2\zeta(3) \right)
+\hat{k}^{4}\left( \frac{13}{16} -\frac{3\pi^{2}}{2}+ \frac{29\pi^{4}}{45} -5\zeta(3) \right) \notag \\
&&~~~~~~~~~~~
+\hat{k}^{6}\left( -\frac{5}{4}-\frac{5\pi^{2}}{6} +\frac{\pi^{4}}{15} -22\zeta(3) \right)
-\frac{5\hat{k}^{8}}{8}
\Biggr) +\mc{O}(\bar{n}^{-5})
\Biggr],
\end{eqnarray}
%
%==================================%
and
%============<Equation>=============%
%
\begin{eqnarray}\label{Imscbb}
{\rm Im}\, \omega_{\pm}&=&-\hat{k}^{2}\Biggl[
1-\frac{1}{\bar{n}}+\frac{1}{\bar{n}^{2}}\left( -1 +\frac{\pi^{2}\hat{k}^{2}}{3} \right)  %\notag \\
%&&\qquad
+\frac{1}{\bar{n}^{3}}\left( -1 -\hat{k}^{2}\left( \frac{4\pi^{2}}{3}+8\zeta(3)\right) -4\hat{k}^{4}\zeta(3) \right) \notag \\
&&\quad
+\frac{1}{\bar{n}^{4}}\Biggl( 
-1 -\hat{k}^{2} \left( \frac{\pi^{2}}{3}-\frac{\pi^{4}}{9} -16\zeta(3) \right)%\notag \\
%&&\qquad\qquad\quad
+\hat{k}^{4} \left( \frac{31\pi^{4}}{45}+36\zeta(3) \right) 
+\frac{4\pi^{4}\hat{k}^{6}}{45}
\Biggr) +\mc{O}(\bar{n}^{-5})
\Biggr].\notag \\
\end{eqnarray}
%
%==================================%

The appearance of the $\zeta$ function in these series will be explained in sec.~\ref{sec:hydro}.

\subsection{Gregory-Laflamme unstable frequencies} 

The AdS/Ricci-flat correspondence of \cite{Caldarelli:2012hy} relates the quasinormal spectrum of $D=\bar{n}+1$ AdS black branes to the spectrum of fluctuations of Ricci-flat black $p$-branes in dimension
\beq
D=\bar{n}+p+3\,.
\eeq
The scalar sector of the latter is known to contain unstable modes \cite{Gregory:1993vy}.

According to \cite{Caldarelli:2012hy}, the map 
requires replacing $\bar{n}\rightarrow -\bar{n}$, so we also have $\hat{k}\rightarrow i\hat{k}$.
By applying this to eqs.~\eqref{Rescbb}, \eqref{Imscbb}, we find the imaginary frequencies $\Omega_\pm=i\omega_{\mp}$ as
%============<Equation>=============%
%
\begin{eqnarray}\label{GLn4}
%\Omega_{\pm} &=& i\omega_{\mp} \notag \\
\Omega_{\pm}&=&
\pm\hat{k}-\hat{k}^{2} -\frac{\hat{k}}{2\bar{n}}\lp\pm 1+2\hat{k}\mp2\hat{k}^{2}\rp \notag \\
&&
+\frac{\hat{k}}{24\bar{n}^{2}}\lp\pm 9 +24\hat{k} \pm12\hat{k}^{2}\mp 8\pi^{2}\hat{k}^{2}+8\pi^{2}\hat{k}^{3}\mp12\hat{k}^{4}\rp \notag \\
&&
+\frac{\hat{k}}{48\bar{n}^{3}}\Bigl[ \mp 15-48\hat{k}\mp 2(27+4\pi^{2}+96\zeta(3))\hat{k}^{2}
+64 (\pi^{2}+6\zeta(3))\hat{k}^{3} \notag \\
&&\qquad\qquad
\mp 12(3+4\pi^{2}-8\zeta(3))\hat{k}^{4}
-192\zeta(3)\hat{k}^{5}\pm 24\hat{k}^{6}
\Bigr] \notag \\
&&
+\frac{\hat{k}}{\bar{n}^{4}} \Biggl[
\pm \frac{35}{128} + \hat{k} \pm\hat{k}^{2}\left( \frac{25}{16}+\frac{3\pi^{2}}{8} -\frac{4\pi^{4}}{45} -2\zeta(3) \right)
+\hat{k}^{3}\left( -\frac{\pi^{2}}{3}+\frac{\pi^{4}}{9}+16\zeta(3) \right) \notag \\
&&\qquad\quad
\pm\hat{k}^{4}\left( \frac{13}{16}-\frac{3\pi^{2}}{2} +\frac{29\pi^{4}}{45}-5\zeta(3) \right)
-\hat{k}^{5} \left( \frac{31\pi^{4}}{45}+36\zeta(3) \right) \notag \\
&&\qquad\quad
\pm\hat{k}^{6} \left( \frac{5}{4} +\frac{5\pi^{2}}{6}-\frac{\pi^{4}}{15} +22\zeta(3) \right)
+\frac{4\pi^{4}\hat{k}^{7}}{45} 
\mp\frac{5\hat{k}^{8}}{8}
\Biggr] +\mc{O}(\bar{n}^{-5}).
\end{eqnarray}
%
%==================================% 
$\Omega_{+}$ is the unstable mode on the black brane. 
We have also obtained this same result by solving directly the perturbations of the Ricci-flat black brane, thus managing to extend the calculation in \cite{Emparan:2013moa} to two higher orders.

\begin{figure}[th]
 \begin{center}
  \includegraphics[width=.6\textwidth]{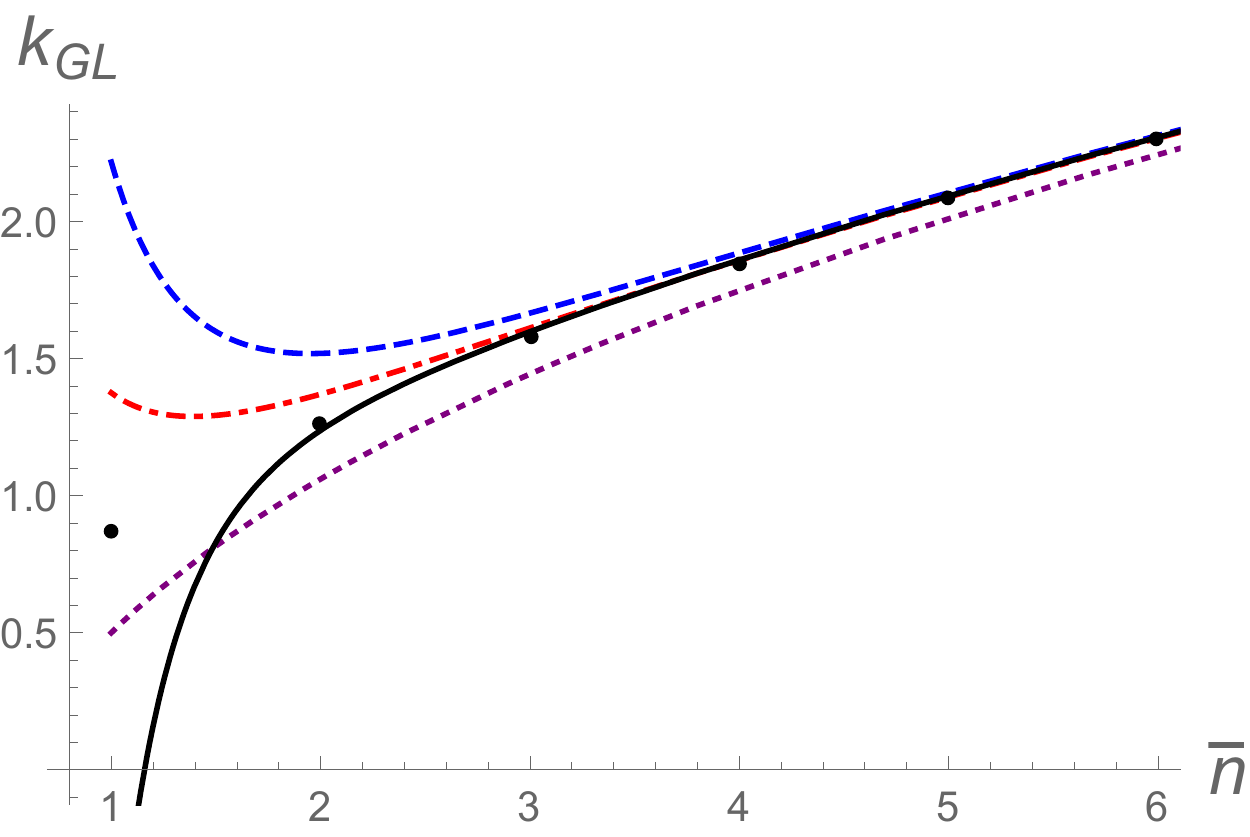}
     \end{center}
% \vspace{-5mm}
 \caption{\small Wavenumber $k_{GL}$ of the marginally stable mode of black $p$-branes as a function of $\bar{n}=D-p-3$. We plot successive approximations to eq.~\eqref{thres}: to $1/\bar{n}$ (purple dotted line); to $1/\bar{n}^2$ (red dash-dotted line); to $1/\bar{n}^3$ (blue dashed line); to $1/\bar{n}^4$ (black solid line). Dots are numerical values from \cite{Asnin:2007rw}. Units are $r_0=1$. }
 \label{fig:kGL}
\end{figure}

Using \eqref{GLn4} we can find the critical wavenumber at the threshold of the instability, \ie\ $k_{\text{GL}}=\hat{k}_{\text{GL}}\sqrt{\bar{n}}$ such that $\Omega_{+}(\hat{k}_{\text{GL}})=O(\bar{n}^{-5})$. We find 
\beq\label{thres}
k_{GL}=\sqrt{\bar{n}}\lp 1-\frac{1}{2\bar{n}}+\frac{7}{8\bar{n}^2}+\lp2\zeta(3)-\frac{25}{16}\rp\frac{1}{\bar{n}^3}+\lp\frac{363}{128}-5\zeta(3)\rp\frac{1}{\bar{n}^4}+\mc{O}(\bar{n}^{-5})\rp\,.
\eeq
In fig.~\ref{fig:kGL} we compare this result to the values found in \cite{Asnin:2007rw} from the numerical solution of the problem. For $\bar{n}=2$, \eg\ a black string in $D=6$, the numerical value is $k_{GL}=1.269$, while \eqref{thres} gives $k_{GL}=1.238$, which is off by $2.4\%$.

\section{Non-perturbative breakdown of decoupling}\label{sec:asym}

We now argue that the $1/D$ expansion for decoupling modes does not converge but is instead only asymptotic, since effects non-perturbative in $1/D$ arise from the breakdown of decoupling at finite $D$. 

Considering for simplicity static perturbations of a Ricci-flat black brane, the solution in the far zone is, up to normalization, given by a modified Bessel function,
\beq
\Psi(r)=\sqrt{r}K_{\frac{\bar{n}+2}{2}} (kr)\,.
\eeq 
The Bessel function at large order has the \textit{asymptotic} expansion 
\beq
K_\nu(x) \sim x^{-\nu}\sum_{i\geq 0}\nu^{-i}a_i(x)\qquad (\nu\gg 1)
\eeq
where $a_i$ is a polynomial in $x$, behaving like $a_i\sim x^i$ so in the overlap zone we have, schematically,
\beq\label{psiexp}
\Psi(r)\sim \frac1{\sqrt{\sR}}\sum_{i\geq 0}\lp\frac{k r_0}{\bar{n}} \rp^i \sR^{i/\bar{n}}\,.
\eeq
We see that 
\beq\label{range}
kr_0\ll \bar{n}\,,
\eeq 
is required for $\Psi(r)$ to satisfy the condition \eqref{decoup} that the mode decouples. This is a limit on the range of momenta to which the large $\bar{n}$ expansion for black branes is applicable, and it
is valid at all orders in the $1/\bar{n}$ expansion. This bound is indeed consistent with the fact that in eqs.~\eqref{vbb}, \eqref{Rescbb}, \eqref{Imscbb} each additional order in $1/\bar{n}$ brings in an additional power $\hat{k}^{2}$.

However, large orders in the expansion \eqref{psiexp}, with $i=\bar{n} +\mc{O}(1)$, give a behavior $\Psi\sim \sqrt{\sR}$ which violates the decoupling condition \eqref{decoup}. This breakdown is non-perturbative in $1/\bar{n}$, and gives a value of
\beq\label{npk}
\textrm{non-perturbative corrections}=\mc{O}\lp\lp \frac{k r_0}{\bar{n} }\rp^{\bar{n}}\rp\,.
\eeq
Generically, for planar horizons (Ricci-flat and AdS black branes) we have $k r_0\sim \sqrt{\bar{n}}$ so we expect
\beq\label{npbranes}
\textrm{non-perturbative corrections}=\mc{O}\lp \bar{n}^{-\bar{n}/2}\rp\qquad (K=0)\,,
\eeq
This implies that, for $\bar{n}$ sufficiently large, the perturbative expansion is reliable only for orders $m$ such that
\beq\label{orderbound}
m\lesssim\frac{\bar{n}}{2}+\mc{O}(1)\,.
\eeq
This rough estimate agrees fairly well with the fact (not shown here) that the inclusion of the $1/\bar{n}^4$ correction in eq.~\eqref{GLn4} gives a better overall fit to numerical calculations of the curve $\Omega_+(k)$ only when $\bar{n}\gtrsim 8$.\footnote{However, the best approximation to $k_{GL}$ (although not to $\Omega_+(k)$ overall) for $\bar{n}=2,3$ is the fourth-order one, as fig.~\ref{fig:kGL} shows. This might be a fortunate accident at low $\bar{n}$, since for $4\lesssim \bar{n}\lesssim 8$ the second-order result for $k_{GL}$ is (slightly) better than the fourth-order one. The third-order correction does not improve on the second-order result until $\bar{n}\gtrsim 12$.}
Note however that \eqref{npk} implies that the accuracy at small $k$ can be considerably better than indicated by \eqref{orderbound}. We elaborate this point in the next section.

For spherical black holes, a similar argument applies replacing $k r_0 \to \ell =\mc{O}(1)$, so the expected corrections in this case are
\beq\label{npbhs}
\textrm{non-perturbative corrections}=\mc{O}\lp n^{-n}\rp\qquad (K=1)\,.
\eeq 
These are smaller than \eqref{npbranes}, which agrees well with the good accuracy found at low $n$ in \cite{Emparan:2014aba} for the computation of decoupled quasinormal frequencies of Schwarzschild black holes. 

To finish, let us mention that the numerical calculations for AdS black branes in \cite{Kovtun:2005ev} of the sound (scalar) frequency in $D=5$ suggest that at large momenta, ${\rm Re}\, \omega_{+}\to k$ and that $|{\rm Im}\, \omega_{\pm}|$ decreases.
This is not visible in our results, so presumably these effects are non-perturbative in $1/\bar{n}$.

\section{Comparison to hydrodynamics}\label{sec:hydro}

Although the $1/D$ series of quasinormal frequencies is not convergent, certain terms in them \textit{can} be resummed to all orders in $1/D$. This is the case at least for vector and scalar frequencies when expanded at low momenta. Indeed these terms have been computed exactly in $D$ in the hydrodynamic expansion \cite{Bhattacharyya:2008mz}.\footnote{Of course the hydrodynamic expansion is not convergent itself \cite{Heller:2013fn}.} 
The vector (shear) mode frequency for an AdS black brane in $D=\bar n+1$ dimensions ($\bar{n}\geq 3$) is\footnote{In order to put the results of \cite{Bhattacharyya:2008mz} in this form, we use the identity $H_m=m^{-1}+H_{m-1}$.}
\beq\label{omshear}
\omega=-i\frac{k^2}{\bar{n}}\lp 1+\frac{k^2}{\bar{n}^2}H_{2/\bar{n}}\rp+\mc{O}(k^6)
\eeq
where $H_m$ is the $m$-th `harmonic number'.
The scalar (sound) mode frequency is
\beq\label{omsound}
\omega_\pm=\pm\frac{k}{\sqrt{\bar{n}-1}}-i\frac{\bar{n}-2}{\bar{n}(\bar{n}-1)}k^2\pm \frac{(\bar{n}-2) \left( 1+H_{2/\bar{n}}\right)}{\bar{n}^2(\bar{n}-1)^{3/2} }k^3+
\mc{O}(k^4)\,.
\eeq
These results can be expanded at large $\bar{n}$ using 
\beq\label{serH}
H_{2/\bar{n}}=-\sum_{j=1}^\infty \zeta(j+1)\lp -\frac{2}{\bar{n}}\rp^{j}\,,
\eeq
which is a convergent series for $|\bar{n}|>2$, and then we find perfect agreement with our large $\bar{n}$ calculations in eqs.~\eqref{vbb}, \eqref{Rescbb}, \eqref{Imscbb}. It is amusing to note that the appearance of the function $H_{2/\bar{n}}$ in these dispersion relations could have been obtained by resumming (using \eqref{serH}) the natural guess for the entire $1/\bar{n}$ series from the terms present in \eqref{vbb}.\footnote{A naive attempt at resumming the highest powers of $k$ in this series gives a result that does not agree well with the large-$k$ behavior in \cite{Kovtun:2005ev}.}

Since the temperature of the black brane (with $r_0=1$) is $T\propto \bar{n}$, the range of applicability of the hydrodynamic expansion, namely $k\ll T$, coincides parametrically with the range of momenta to which the large $\bar{n}$ results apply \eqref{range}. In this view, the large  $\bar{n}$ expansion does not afford a larger regime of applicability than the hydrodynamic expansion. The main advantage of the large  $\bar{n}$ expansion is that it allows a simpler computation of higher powers of the momenta (in our case, up to $k^9$ for the sound mode, and $k^8$ for the shear mode), some of which can be relevant (barring non-perturbative effects) at relatively low dimensions. These results also contain information about the transport coefficients at higher orders in the gradient expansion.

\section{Conclusions}\label{sec:conc}

The large $D$ expansion of black hole perturbations provides a natural means to isolate the sector of quasinormal modes that usually contains the most interesting dynamics of the black hole. This is the decoupled spectrum. In this article we have shown that the large $D$ expansion allows efficient calculation of these modes giving unified, analytic expressions for all the neutral, static black holes whose solutions are known in closed form,\footnote{We have not considered static black rings nor other static blackfolds  in de Sitter space \cite{Caldarelli:2008pz,Armas:2010hz}, which are not known in closed exact form.} up to relatively high orders in the expansion.

These results have allowed us to examine the lack of convergence of the series. This is not surprising, since the property of decoupling holds to all perturbative orders in $1/D$ but is obviously absent at finite values of $D$. However, it is less clear whether calculations for non-decoupling modes are equally affected by these limitations.

\section*{Acknowledgments}

Part of this work was done during the workshop ``Holographic vistas on Gravity and Strings'' YITP-T-14-1 at the Yukawa Institute for Theoretical Physics, Kyoto University, whose kind hospitality we acknowledge. While there, we had very useful discussions with Vitor Cardoso, \'Oscar Dias and Paolo Pani.
Work supported by FPA2010-20807-C02-02, FPA2013-46570-C2-2-P, AGAUR 2009-SGR-168 and CPAN CSD2007-00042 Consolider-Ingenio 2010. KT was supported by a JSPS grant for research abroad, and by JSPS Grant-in-Aid for Scientific Research No.26-3387.

%\newpage

\addcontentsline{toc}{section}{Appendices}
%\addtocontents{toc}{\protect\setcounter{tocdepth}{0}}
\appendix

\section{Scalar-type perturbations}\label{app:scpert}

For scalar-type perturbations in the `Regge-Wheeler' formulation of \eqref{waveeq}, the minimum of the potential is at $\sR\sim D$. Although this is within the near-horizon zone, it requires one to deal separately with two (overlapping) regions, one with $\sR=\mc{O}(1)$ and another with $\sR=\mc{O}(D)$. Although this can be done \cite{Emparan:2014aba}, it is simpler to work instead with a different set of gauge invariant variables,
$X$, $Y$ and $Z$, in terms of which the equations are \cite{Kodama:2003jz}
%============<Equation>=============%
%
\begin{eqnarray}
&&
X'(r) = \frac{D-4}{r}X(r) +\left( \frac{f'(r)}{f(r)} -\frac{2}{r} \right) Y(r) 
+\left( \frac{q^{2}}{r^{2}f(r)}-\frac{\omega^{2}}{f(r)^{2}} \right) Z(r)\,, \label{Xeq}
\end{eqnarray}
\begin{eqnarray}
Y'(r) =\frac{f'(r)}{2f(r)}(X(r)-Y(r)) +\frac{\omega^{2}}{f(r)^{2}}Z(r), \label{Yeq}
\end{eqnarray}
\begin{eqnarray}
Z'(r) = X(r) \label{Zeq}
\end{eqnarray}
together with the consistency condition
\begin{eqnarray}
&&
\Bigl[ 
\omega^{2}r^{2} +K \lambda r^{2}+\frac{1}{2\sR}\left( (D-2)(D-3)K-(D-1)(D-2)\lambda r^{2}-\frac{(D-1)(D-3)}{2\sR} \right)
\Bigr]X(r) \notag \\
&&~~~
+\Bigl[ 
\omega^{2}r^{2} -q^{2} f(r)+(D-2)K^{2}-(D-3)K\lambda r^{2} -\frac                                                                               {2(D-2)}{\sR} + \frac{(D-1)^{2}}{4\sR^{2}}
\Bigr]Y(r) \notag \\
&&~~~
-\frac{1}{r}\Bigl[
(D-2)\omega^{2}r^{2} -q^{2}\left(\frac{1}{2\sR}-\lambda r^{2}\right)
\Bigr]Z(r) =0. \label{EQ}
\end{eqnarray}
%
%==================================%

If we introduce two variables $P(\sR)$ and $Q(\sR)$ as
%============<Equation>=============%
%
\begin{eqnarray}
X(\sR)=P(\sR)+\frac{\sR}{\sR-\sR_0}Q(\sR),~~
Y(\sR)=P(\sR)-\frac{\sR}{\sR-\sR_0}Q(\sR),
\end{eqnarray}
%
%==================================%
then the perturbation equations in the near-horizon zone decouple in these variables
at each order in $1/D$. Then they can be solved to yield decoupled quasinormal modes without needing to split into two zones.

\section{Another perturbative formulation for AdS black branes}\label{app:KS}

The perturbation equations for AdS black branes can be analyzed in a different manner, as described in \cite{Kovtun:2005ev}.

We write the background metric as
%============<Equation>=============%
%
\begin{eqnarray}
ds^{2}=r^{2}(-\hat{f}(r)dt^{2} +\delta_{ab}dx^{a}dx^{b}) + \frac{dr^{2}}{r^{2}\hat{f}(r)},
\end{eqnarray}
%
%==================================%
where $\hat{f}(r)=1-r^{-(D-1)}$ and the coordinates $x^{a}$ span $\mathbb{R}^{D-2}$. For linearized perturbations we can always choose a coordinate $z$ aligned with the momentum of the perturbation, $k_a x^a=k z$, so the dependence on $x^{a}$ takes the form $\sim e^{ikz}$. Then perturbations can be decomposed into scalars, vectors and tensors  with respect to the little group of $k^a$, $SO(D-3)$.
In the following we write 
\beq
x^{a}=(z,x^{i})\qquad (i=1,\dots, D-3).
\eeq   
It is easy to check the equivalence between these perturbations and those of \cite{Kodama:2003jz}. For example, the metric perturbation for a $\mathbb{R}^{D-2}$-scalar perturbation in \cite{Kovtun:2005ev} is, in momentum space,
%============<Equation>=============%
%
\begin{eqnarray}
h^{S_{T}}_{ab}=S_{T}(t,r)\left( k_{a}k_{b}-\frac{k^{2}\delta_{ab}}{D-2}  \right)\,.
\end{eqnarray}
%
%==================================%  
When we align $k_{a}$ with $k_{z}$ the only non-vanishing component of $h^{S_{T}}_{ab}$ is $h^{S_{T}}_{zz}$, which is scalar-type with respect to $SO(D-3)$. One can find similar relations for other variables in the two decompositions.

The perturbation equations can be written in terms of master variables $Z_{s}$. For tensors ($s=2$)  and vectors ($s=1$) these are
%============<Equation>=============%
%
\begin{eqnarray}
&&
Z_{s=2}\tau_{ij}=h_{ij}, \\
&&
Z_{s=1}\partial_{i}=k h_{ti}+\omega h_{zi},
\end{eqnarray}
while for scalars ($s=0$)
\begin{eqnarray}
&&Z_{s=0}=4\omega k h_{tz} +2\omega^{2} h_{zz} \notag \\
&&~~~~~~
+\frac{1}{D-3}\left( k^{2}((D-1)-(D-3)\hat{f}(r))-2\omega^{2} \right) \delta^{ij}h_{ij} +2k^{2}h_{tt},
\end{eqnarray}
%
%==================================%
Here $\tau_{ij}$ is a symmetric traceless tensor. 
Note that since we choose $Z_{s}\sim e^{ikz}$ there are no scalar-derived tensor or vector perturbations, nor vector-derived tensor perturbations. 

The perturbation equation for tensors $Z_{s=2}$ is
%============<Equation>=============%
%
\begin{eqnarray}
Z''_{s=2} +\frac{D-1+\hat{f}(r)}{r \hat{f}(r)}Z'_{s=2} +\frac{\omega^{2} -k^{2}\hat{f}(r)}{r^{4}\hat{f}(r)^{2}}Z_{s=2}=0,
\end{eqnarray}
%
%==================================%
where the prime is derivative with respect to $r$. 

The equations for the vectors and scalars are
%============<Equation>=============%
%
\begin{eqnarray}
Z''_{s=1}+\left( \frac{D}{r}+\frac{\hat{f}'(r)\omega^{2}}{\hat{f}(r)(\omega^{2}-q^{2}\hat{f}(r))} \right)Z'_{s=1}
+\frac{\omega^{2}-k^{2}\hat{f}(r)}{r^{4}\hat{f}(r)^{2}}Z_{s=1}=0
\end{eqnarray}
and
\begin{eqnarray}
Z''_{s=0}+\frac{Y_{1}k^{2}+Y_{2}\omega^{2}}{r\hat{f}(r)X}Z'_{s=0}
+\frac{Y_{3}k^{2}+Y_{4}k^{4}+2(D-2)\omega^{4}}{r^{4}\hat{f}(r)^{2}X}Z_{s=1}=0,
\end{eqnarray}
%
%==================================%
where
%============<Equation>=============%
%
\begin{eqnarray}
\begin{aligned}
&X=2(D-2)\omega^{2} -((D-1)+(D-3)\hat{f}(r))k^{2}, \\
&Y_{1}=-(2D-1)(D-3)\hat{f}(r)^{2}-(D-1)(D-1- (D-4)\hat{f}(r)), \\
&Y_{2}=2(D-2)(D-1+\hat{f}(r)), \\
&Y_{3}=-(D-3)r^{4}(\hat{f}'(r))^{2}\hat{f}(r)-((3D-7)\hat{f}(r)+(D-1))\omega^{2},\\
&Y_{4}=((D-1)+(D-3)\hat{f}(r))\hat{f}(r).
\end{aligned}
\label{KSquan}
\end{eqnarray}
%
%==================================%
These equations can now be solved perturbatively in $1/D$ in the near-horizon zone in the usual manner, yielding the results in sec.~\ref{sec:adsb}.

\section{$\omega_{(3)}$ for scalar modes}\label{app:om3sc}

\beqa
\textrm{Re}\,\omega_{(3)}&=&\frac{1}{48\sR_0^3\left(\frac{\hat{q}^2}{\sR_0}-K^2\right)^{5/2}}
\Biggl[24 \hat{q}^{12}\sR_0^3\nn\\
&&-4 \hat{q}^{10}\sR_0^2 \left(2 K\sR_0 \left(9 \ln \sR_0 +24 \zeta (3)+2 \pi ^2+27\right)-3 \left(8 \ln \sR_0 -8 \zeta (3)+4 \pi ^2+21\right)\right)\nn\\
&&
+2 \hat{q}^8\sR_0 \biggl(8 K^3\sR_0^3 \left(24 \zeta (3)+\pi ^2\right)
\nn\\
&&\qquad\qquad +12 K^2\sR_0^2 \left(3 \ln ^2\sR_0 +\left(21+2 \pi ^2\right) \ln \sR_0  +48 \zeta (3)+2 \pi ^2+21\right)
\nn\\
&&\qquad\qquad-4 K\sR_0 \left(18 \ln ^2\sR_0 +3 \left(47+2 \pi ^2\right) \ln \sR_0 -12 \zeta (3)+31 \pi ^2+159\right)
\nn\\
&&\qquad\qquad+48 \ln ^2\sR_0 -8 \left(2 \pi ^2-45\right) \ln \sR_0 -96 \zeta (3)-60 \pi ^2+477\biggr)\nn\\
&&-\hat{q}^6 \biggl(15+192 K^5\sR_0^5 \zeta (3)+16 K^4\sR_0^4 \left(\left(6+5 \pi ^2\right) \ln \sR_0 +2 \left(63 \zeta (3)+5 \pi ^2\right)\right)\nn\\
&&\qquad+24 K^3\sR_0^3 \left(\ln ^3\sR_0 +17 \ln ^2\sR_0 +\left(23+2 \pi ^2\right) \ln \sR_0 +72 \zeta (3)-7 \pi ^2+1\right)\nn\\
&&\qquad-4 K^2\sR_0^2 \left(12 \ln ^3\sR_0 +195 \ln ^2\sR_0 +8 \left(60+7 \pi ^2\right) \ln \sR_0 +192 \zeta (3)+184 \pi ^2+57\right)\nn\\
&&\qquad-2 K\sR_0 \left(-16 \ln ^3\sR_0 -252 \ln ^2\sR_0 +\left(16 \pi ^2-969\right) \ln \sR_0 +96 \zeta (3)+52 \pi ^2-819\right)\biggr)\nn\\
&&
+2 K^2 \hat{q}^4\sR_0 \biggl(16 K^4\sR_0^4 \left(\pi ^2 \ln \sR_0 +30 \zeta (3)+2 \pi ^2\right)\nn\\
&&\qquad\qquad+8 K^3\sR_0^3 \left(18 \ln ^2\sR_0 +12 \left(3+\pi ^2\right) \ln \sR_0 +198 \zeta (3)+17 \pi ^2\right)\nn\\
&&\qquad\qquad-12 K^2\sR_0^2 \left(-\ln ^3\sR_0 +11 \ln ^2\sR_0 +\left(63+10 \pi ^2\right) \ln \sR_0 +8 \zeta (3)+37 \pi ^2+2\right)\nn\\
&&\qquad\qquad-4 K\sR_0 \left(4 \ln ^3\sR_0 +18 \ln ^2\sR_0 +\left(22 \pi ^2-135\right) \ln \sR_0 +96 \zeta (3)+61 \pi ^2-234\right)\nn\\
&&\qquad\qquad+16 \ln ^3\sR_0 +180 \ln ^2\sR_0 +501 \ln \sR_0 +324\biggr)\nn\\
&&-4 K^4 \hat{q}^2\sR_0^2 \biggl(24 K^3\sR_0^3 \left(\ln ^2\sR_0 +\left(2+\pi ^2\right) \ln \sR_0 +2 \left(8 \zeta (3)+\pi ^2\right)\right)\nn\\
&&\qquad\qquad+4 K^2\sR_0^2 \left(6 \ln ^2\sR_0 +\left(\pi ^2-12\right) \ln \sR_0 -9 \left(\pi ^2-8 \zeta (3)\right)\right)\nn\\
&&\qquad\qquad-2 K\sR_0 \left(-\ln ^3\sR_0 +51 \ln ^2\sR_0 +6 \left(13+5 \pi ^2\right) \ln \sR_0 +10 \left(12 \zeta (3)+8 \pi ^2-9\right)\right)\nn\\
&&\qquad\qquad+4 \ln ^3\sR_0 +99 \ln ^2\sR_0 +354 \ln \sR_0 +234\biggr)\nn\\
&&+8 K^6\sR_0^3 \biggl(4 K^2\sR_0^2 \left(3 \ln ^2\sR_0 +2 \left(3+\pi ^2\right) \ln \sR_0 +4 \left(6 \zeta (3)+\pi ^2\right)\right)\nn\\
&&\qquad\qquad-4 K\sR_0 \left(6 \ln ^2\sR_0 +3 \left(6+\pi ^2\right) \ln \sR_0 +12 \zeta (3)+8 \pi ^2\right)\nn\\
&&\qquad\qquad+\ln ^3\sR_0 +18 \ln ^2\sR_0 +72 \ln \sR_0 +48\biggr)
\Biggr]\,,
\eeqa

\beqa
\textrm{Im}\,\omega_{(3)}&=&4 \hat{q}^6\sR_0^2 \zeta (3)\nn\\
&&
+\frac{\hat{q}^4\sR_0}{3}  \biggl(3 \left(\pi ^2 \ln \sR_0 +8 \zeta (3)+4 \pi ^2\right)-2 K\sR_0 \left(\pi ^2 \ln \sR_0 +30 \zeta (3)+2 \pi ^2\right)\biggr)\nn\\
&&
+\frac{\hat{q}^2}{6} \biggl(12 K^2\sR_0^2 \left(\ln ^2\sR_0 +\left(2+\pi ^2\right) \ln \sR_0 +2 \left(8 \zeta (3)+\pi ^2\right)\right)\nn\\
&&\qquad-2 K\sR_0 \left(12 \ln ^2\sR_0 +\left(42+9 \pi ^2\right) \ln \sR_0 +48 \zeta (3)+29 \pi ^2\right)\nn\\
&&\qquad+\ln ^3\sR_0 +21 \ln ^2\sR_0 +102 \ln \sR_0 +90\biggr)\nn\\
&&
-\frac{K}{6}  \biggl(4 K^2\sR_0^2 \left(3 \ln ^2\sR_0 +2 \left(3+\pi ^2\right) \ln \sR_0 +4 \left(6 \zeta (3)+\pi ^2\right)\right)\nn\\
&&\qquad-4 K\sR_0 \left(6 \ln ^2\sR_0 +3 \left(6+\pi ^2\right) \ln \sR_0 +12 \zeta (3)+8 \pi ^2\right)\nn\\
&&\qquad+\ln ^3\sR_0 +18 \ln ^2\sR_0 +72 \ln \sR_0 +48\biggr)\,.
\eeqa

\section{Results for (A)dS-Schwarzschild black holes}\label{app:adsbh}

In this case $K=1$ and
%\beq \sR_0=\lp 1+\frac{r_0^2}{L^2}\rp^{-1} \eeq with $L$ the AdS radius. T
the eigenvalue $\hat q^2$ must be expanded in $1/n$ like in \eqref{eigen}. All the dependence on the cosmological constant is included in 
\beq
\sR_0=\frac{1}{1-\lambda}=\frac{L^2}{L^2+r_0^2}\,,
\eeq
where in the last expression we employ $\lambda=-r_0^2/L^2$, which is convenient for AdS with radius $L$.

\paragraph{Vector-type}

%============<Equation>=============%
%
\begin{eqnarray}
&&\omega r_{0} = 
-i(\ell-1)\Biggl[ 1 +\frac{1}{n}(\ell-1-\ln{\sR_0} ) \notag \\
&& 
\qquad +\frac{1}{n^{2}} \biggl( 
\left( \frac{\pi^{2} \sR_0}{3}-2\right)(\ell-1 ) - (\ell-3+2 \sR_0)\ln{\sR_0} 
+\frac{1}{2}(\ln{\sR_0})^{2}
\biggr) \notag \\
&&
\qquad +\frac{1}{6n^3} \biggl(
4(\ell-1)\left(\pi^{2} \sR_0 (2\sR_0 +\ell-3) -6\sR_0^{2}(\ell-1)\zeta(3)
-6 \sR_0  \zeta (3)+6\right) \notag \\
&&
\qquad\quad +2\ln{\sR_0}\left( -2\left(6+\pi^{2}\right)\sR_0^{2}+3\left( 10+\pi^{2}\right)\sR_0 
+\left( 2\pi^{2}\sR_0^{2}-3 \left(2+\pi^{2}\right)\sR_0 +12\right) \ell-24\right) \notag \\
&&
\qquad\quad +3(\ln{\sR_0})^{2}\left(-4 \sR_0^{2}+8 \sR_0 +\ell -5\right)-(\ln{\sR_0})^{3}\biggl)  +\mc{O}(n^{-4})\Biggr],
\end{eqnarray}

\paragraph{Scalar-type}
%============<Equation>=============%
%
\begin{eqnarray}
&&
{\rm Re}\, \omega_{\pm}r_{0} =\pm\sqrt{\frac{\ell-\sR_0}{\sR_0}}\Biggl[
1+\frac{\ell-1}{2(\ell-\sR_0)n}\left( (2 \sR_0+1)\ell-4 \sR_0 -2 \sR_0 \ln{\sR_0}\right)\notag \\
&&
+\frac{\ell-1}{24(\ell-\sR_0)^{2}n^{2}}\Bigl(
   16 \sR_0^{2} \left(\pi^{2} \sR_0 -6\right)-3 \left(4 \sR_0^{2}-12 \sR_0 +1\right) \ell^{3} \notag \\
&&~~
+\left(4\left(3+2 \pi^{2}\right) \sR_0^{2}-\left(132-8 \pi^{2}\right) \sR_0 -9\right) \ell^{2}-
4\sR_0 \left(2 \pi^{2} \sR_0^{2}+6 \left(\pi^{2}-3\right) \sR_0 -33\right) \ell \notag \\
&&~~
+12\sR_0 \ln{\sR_0} \left(4 (\sR_0 -2) \sR_0 +(2 \sR_0 -5) \ell^{2}+(11-4 \sR_0 ) \ell\right)
-12 \sR_0  (\ln{\sR_0})^{2} (\sR_0 +(\sR_0 -2) \ell)\Bigr) \notag \\
&&
+\frac{\ell-1}{48(\ell-\sR_0)^{3}n^{3}}\left( \Bigl(3 \left(8 \sR_0^{3}-20 \sR_0^{2}+6 \sR_0 +1\right)
\ell^{5}-2 \ell^{4} (-12 \left(\pi^{2}-15\right) \sR_0\right. \notag \\
&&~~~~~~
+8 \sR_0^{3} \left(12 \zeta(3)+6+\pi^{2}\right)
-12\sR_0^{2} \left(-4 \zeta(3)+15+3 \pi^{2}\right)-3) \notag \\
&&~~~~
+\ell^{3}(-4 \left(51+64 \pi^{2}\right) \sR_0^{2}+16 \sR_0^{4} \left(24 \zeta (3)
+\pi^{2}\right)\notag \\
&&~~~~~~
-24 \sR_0^{3}\left(-40 \zeta(3)-1+\pi^{2}\right)-2\sR_0\left(96 \zeta (3)-447+52 \pi^{2}\right)+15)\notag \\
&&~~~~
-8 \sR_0  \ell^{2}(24 \sR_0^{4} \zeta (3)+2 \sR_0^{3} \left(102 \zeta (3)+7
\pi^{2}\right)+\sR_0^{2} \left(96 \zeta (3)-6-31 \pi^{2}\right) \notag \\
&&~~~~~~
+\sR_0    \left(-96 \zeta (3)+114-61 \pi ^2\right)+81)  \notag \\
&&~~~~
+8\sR_0^{2} \ell \left(8  \sR_0^{3} \left(12 \zeta (3)+\pi^{2}\right)+2 \sR_0^{2} 
\left(96 \zeta (3)+7 \pi^{2}\right)\right.\notag \\
&&~~~~~~\left.
-2 \sR_0  \left(60 \zeta (3)-21+40 \pi^{2}\right)+117\right)\notag \\
&&~~~~
-128 \sR_0^{3} \left(\pi^{2} (\sR_0 -2) \sR_0 +6 \sR_0^{2} \zeta (3)-3 \sR_0\zeta (3)
+3\right)\Bigr)\notag \\
&&~~~~
-2 \sR_0 \ln{\sR_0}(16 \sR_0^{2} \left(2 \left(3+\pi^{2}\right) \sR_0^{2}-3 \left(6+\pi^{2}\right) \sR_0 
+18\right)+\left(36 \sR_0^{2}-84 \sR_0 +69\right) \ell^{4} \notag \\
&&~~~~
-2 \left(4 \pi^{2} \left(3 \sR_0^{2}-3 \sR_0-2\right)+3 \left(24 \sR_0^{2}-66 \sR_0 +67\right)\right) \ell^{3} \notag \\
&&~~~~
+\left(8 \left(6+5 \pi^{2}\right) \sR_0^{3}-36 \sR_0^{2}-4 \left(45+22 \pi^{2}\right)\sR_0 +501\right) \ell^{2} \notag \\
&&~~~~
-4 \sR_0  \left(4 \pi^{2} \sR_0^{3}+2 \left(24+7 \pi^{2}\right) \sR_0^{2}-30 \left(5+\pi^{2}\right) \sR_0 
+177\right)\ell) \notag \\
&&~~~~
+12 \sR_0 (\ln{\sR_0})^{2}(-4 \sR_0^{2} \left(2 \sR_0^{2}-4 \sR_0+3\right)+\left(6 \sR_0^{2}
-13 \sR_0 +10\right) \ell^{3}+\left(-24 \sR_0^{2}+44\sR_0 -30\right) \ell^{2} \notag \\
&&~~~~
+\sR_0  \left(24 \sR_0^{2}-46 \sR_0 +33\right)\ell) - 8 \sR_0  (\ln{\sR_0})^{3} \left(\sR_0^{2}
+\left(3 \sR_0^{2}-6 \sR_0 +4\right) \ell^{2}-2 \sR_0  \ell\right)
 \Bigr)
\nn\\&& +\mc{O}(n^{-4})\Biggr],
\end{eqnarray}
%
%==================================% 
and
%============<Equation>=============%
%
\begin{eqnarray}
&&
{\rm Im}\, \omega_{\pm}r_{0} = -i(\ell-1)\Biggl[
1+ \frac{1}{n}(\ell-2-\ln{\sR_0}) \notag \\
&&~~~~
+\frac{1}{n^2}\left(4-3 \ell+(\ell-2)\frac{\pi^{2}\sR_0}{3}
-(2\sR_0 +\ell -4)\ln{\sR_0}
+\frac{(\ln{\sR_0})^{2}}{2} \right) \notag \\
&&~~~~
+\frac{1}{n^3}\left(\left(2 \sR_0 \ell^{2}\left(\frac{\pi^{2}}{3}-2\sR_0 \zeta (3)\right)
+ \ell\left(8 \sR_0(2\sR_0 -1) \zeta (3)+7-\frac{\pi^{2}}{3} \sR_0  (13-4 \sR_0)\right)\right.\right. \notag \\
&&~~~~~~\left. 
+8\left(\frac{\pi^{2}}{3} (2-\sR_0) \sR_0 -\sR_0(2 \sR_0-1) \zeta (3)-1\right)\right) \notag \\
&&~~~~~~
-\ln{\sR_0} \left(4 \left(1+\frac{\pi^{2}}{3} \right) \sR_0^{2}-2 \left(6+\pi^{2}\right) \sR_0 
-\left(\frac{2\sR_0^{2}\pi^{2}}{3} -\left(2+\pi^{2}\right) \sR_0 +5\right) \ell +12\right) \notag \\
&&~~~~~~\left.
+\frac{1}{2} (\ln{\sR_0})^{2} \left(-4 \sR_0^{2}+8 \sR_0 +\ell -6\right)-\frac{(\ln{\sR_0})^{3}}{6} \right)+\mc{O}(n^{-4})
\Biggr].
\end{eqnarray}

%%%%%%%%%%%%%%%%%%%%%%%%%%%%%%%%%%%%%%%%%

\end{document}